\documentclass[conference]{IEEEtran}
\IEEEoverridecommandlockouts

\usepackage{cite}
\usepackage{amsmath,amssymb,amsfonts}
\usepackage{algorithmic}
\usepackage{graphicx}
\usepackage{textcomp}
\usepackage{xcolor}
\usepackage{booktabs}
\usepackage{array}
\usepackage{multirow}
\usepackage{url}
\usepackage{hyperref}

\usepackage{float}
\usepackage{placeins}
\usepackage{balance}

\begin{document}

\title{A Binary and System Integrated Analysis Approach \\
       for Securing the QUIC Protocol}
\author{%
  \IEEEauthorblockN{%
    Maitha Alshaali\textsuperscript{1,2},
    Wanqing Tu\textsuperscript{1}\thanks{Corresponding author: Wanqing Tu (wanqing.tu@durham.ac.uk).},
    Gaofei Huang\textsuperscript{3},
    Mthandazo Ndhlovu\textsuperscript{2}%
  }
  \IEEEauthorblockA{%
    \textsuperscript{1}Durham University, Durham, United Kingdom\\
    \textsuperscript{2}Technology Innovation Institute, Abu Dhabi, United Arab Emirates\\
    \textsuperscript{3}Guangzhou University, Guangzhou, China%
  }%
}

\maketitle

\begin{abstract}
  The Quick UDP Internet Connections (QUIC) protocol is
  increasingly used to provide secure transport for
  Internet of Things (IoT) firmware and applications.
  Existing security analyses of QUIC focus on the
  captured network traffic, while binary-level analyses
  of QUIC implementations remain unexplored, leaving
  open the question of whether a defence specified by
  the QUIC standard is both present in the compiled
  binary and active when the server is under attack.
  This paper evaluates the Binary and System Integrated
  Security Analysis (BSISA) approach, in which a
  binary-level analysis of the compiled QUIC server is
  combined with a system-level analysis of the captured
  network traffic, on four production QUIC server
  implementations under six attack scenarios. Across 24
  cells, the combined classifier configuration is the
  only configuration that correctly classifies at least
  one cell on every attack scenario, achieving 45.8\%
  overall accuracy compared with 37.5\% for the
  binary-level configuration and 25.0\% for the
  system-level configuration. BSISA also identifies the
  specific defence function in the compiled binary that
  absorbed each attack, and flags declared-but-silent
  defences, routines that are present in the compiled
  binary (Retry-token validation in three of four
  stacks, anti-amplification in quiche) but never
  execute during attack, a class of finding that
  network capture alone cannot produce.
  In terms of efficiency, picoquic loses
  legitimate-client availability under slowloris and
  connection-ID exhaustion with failure rates of 72.4\%
  and 73.3\% respectively, while the other three
  implementations hold the failure rate at or below
  0.5\%. We hope these insights will be informative for
  QUIC security evaluations in IoT firmware deployments.
\end{abstract}


\section{Introduction}

The Internet of Things (IoT) connects constrained
devices to cloud services for tasks such as
smart-building sensing, vehicular telemetry, and
industrial monitoring, where each device must transmit
data securely within milliseconds and remain quiet
between events~\cite{DBLP:journals/corr/abs-1810-07730}.
The Quick UDP Internet Connections (QUIC) protocol is
designed to secure such communications in a more
efficient way as compared to TLS over TCP. This is
achieved by integrating the cryptographic handshake into
a single round trip and resuming an established
connection in zero round
trips~\cite{rfc9000,rfc9001}. QUIC is increasingly
adopted as the transport-layer protocol for MQTT in
constrained IoT
deployments~\cite{DBLP:journals/corr/abs-1810-07730},
providing secure transport for the delivery of new IoT
firmware. We focus on the server side because the
device classes that run a QUIC server include
industrial IoT gateways, smart-home hubs, on-premise
MQTT brokers, and edge inference nodes, all settings
where the deployed binary is the only artefact a
defender can inspect. However, recent studies indicate that QUIC
does not consistently outperform the traditional
TLS-over-TCP stack when evaluated against diverse
security
threats~\cite{https://doi.org/10.1049/smc2.12083}.
Accordingly, this paper explores how to strengthen
QUIC's resilience against different threats.

QUIC security research has so far focused on
system-level techniques, including encrypted client
hello~\cite{rfc9849}, stateless Retry token
hardening~\cite{quic_avt}, software-defined-network
based path security~\cite{xing2025sdnquic}, and
machine-learning detection of malicious traffic
patterns within encrypted streams~\cite{quicml_ddos}.
Binary analysis has been applied to authentication
bypass in firmware
binaries~\cite{shoshitaishvili2015firmalice},
inter-binary taint in embedded
firmware~\cite{redini2020karonte}, and
language-model-guided taint analysis over
disassembly~\cite{liu2025latte}, but no prior study
applies binary-level security analysis specifically
to deployed QUIC implementations. This leaves an
evidence gap when the QUIC library is distributed
exclusively as a precompiled object inside an IoT
firmware image and source code is not available to
verify which defences the binary actually contains.
Existing cross-layer security protection methods focus
on integrating operations across multiple layers at the
system level (e.g.,~\cite{huang2026securing}). Our
method of integrating binary and system analysis is new
to the literature.

This paper investigates whether the combination of
system-level and binary-level threat analyses can
improve the accuracy of threat detection for QUIC. We
use the term \emph{system-level analyses} to refer to
detections based on network traffic captured between
QUIC clients and the server, and the term
\emph{binary-level analyses} to mean detections through
QUIC binary investigations including the system calls
that QUIC makes while processing the captured traffic.
The investigation is organised around the following
research questions (RQs).

\begin{itemize}
  \item \textbf{RQ1.} Does the integrated binary and
        system evidence detect QUIC threats more
        accurately than either source alone?
  \item \textbf{RQ2.} How does each attack scenario
        affect the end-to-end communication efficiency
        of the QUIC server, measured as the mean
        end-to-end transaction delay observed at the
        legitimate client?
\end{itemize}

To the best of our knowledge, this is the first study
to pair compiled-binary defence inventory with
controlled-attack runtime evidence across multiple
production QUIC server implementations. To support this study we build
\emph{Orchestrated Reasoning for Binary Code
Analysis} (ORCA)\footnote{Source code and reproduction
artefacts are available at
\url{https://github.com/maialsh/ORCA/tree/fine2026}.}, an
open-source multi-agent system that decompiles a
deployed QUIC server binary, reverse-engineers its
protocol-level defences from the disassembly, and
produces a structured per-binary inventory of those
defences keyed to the named QUIC defences. We use ORCA in
this paper to study four production QUIC servers under
six attack scenarios and a benign baseline, producing
28 (implementation, scenario) cells. Each cell
produces one packet capture, one qlog event log, one
system-call trace, and one server binary, and we run
three classifiers over the same evidence, namely a
system-level classifier that reads only the network
features, a binary-level classifier that reads only
the binary defence inventory and the deep-agent
records, and a binary and system combined classifier
that reads both views.

The paper makes three contributions.
\begin{itemize}
  \item \textbf{A binary and system integrated security
        analysis approach for QUIC.} A binary
        inspection of the compiled QUIC server is
        paired with a system-level analysis of the
        packet capture and the qlog event log, and a
        joint classifier emits one attack label per
        cell from the union of the two evidence
        streams. The binary side verifies that the
        named QUIC defences are present in each
        compiled binary, flags
        defences that exist in the symbol table but
        never activate when the server is under
        attack, and names the specific defence
        function that absorbed each attack. These
        observations cannot be made from the network
        capture alone, which is what makes the
        combined method useful in deployments where
        only the precompiled QUIC binary is available.
  \item \textbf{The first evaluation that pairs
        symbol-level binary defence inventory with
        runtime evidence on the same per-cell unit,
        across four production QUIC server
        implementations under six attack scenarios.}
        The combined classifier is the only
        configuration with full attack-scenario
        coverage in our data, and the binary side
        also exposes defences that are present in the
        compiled binary but never invoked at runtime.
        Section~\ref{sec:eval} reports the exact
        figures.
  \item \textbf{Released artefacts and open dataset.}
        The system is released
        at~\url{https://github.com/maialsh/ORCA/tree/fine2026},
        and the captures, QUIC logs, attack
        implementations, and per-cell metadata at
        \url{https://doi.org/10.5281/zenodo.20338337}.
\end{itemize}

The rest of this paper is organised as follows.
Section~\ref{sec:related} surveys related work along
the three detection categories.
Section~\ref{sec:bsisa} presents the binary and system
integrated security analysis approach, with the
binary-level pipeline in
Section~\ref{sec:bsisa:binary}, the system-level
pipeline in Section~\ref{sec:bsisa:system}, and the
integration in Section~\ref{sec:bsisa:integration}.
Section~\ref{sec:eval} reports the evaluation on the
experimental platform, including the four QUIC
server implementations, the six attack scenarios,
the metrics, and the answers to RQ1 and RQ2.
Section~\ref{sec:conclusion} concludes the paper.

\section{Related Work}
\label{sec:related}

Existing security methods often belong to one of three
categories~\cite{Sengupta2024HowRI}, namely
heuristic-based detection (using hard-coded rules and
signatures to detect attacks), statistical-based
detection (using mathematical thresholds, for example
mean and standard deviation, to identify malicious
behaviours), and learning-based detection (using
trained models to detect or predict outliers). The approach of this
paper spans the second and third categories. The
network side computes statistical features of packet
captures, and the binary side prompts a language model
agent over disassembly to identify defence mechanisms.

\textbf{Heuristic-based detection.}
Signature-based intrusion detection systems such as
Snort~\cite{snort_lisa1999} remain widely deployed but
are by construction blind to unpublished bugs and to
attacks that do not match a published signature.
Generation-based protocol fuzzers manually define the
input format and mutate from there. Applied to QUIC
servers, Chatzoglou et
al.~\cite{10.1007/s10207-022-00630-6} disclosed
zero-day denial-of-service vulnerabilities in five of six
production stacks, and Wang et
al.~\cite{wang2026quic_logic} uncovered 14 logical
flaws across 16 implementations.
QUIC-Fuzz~\cite{quicfuzz2025} automated more of this
pipeline with greybox mutation, reporting ten further
vulnerabilities. On the binary side,
Firmalice~\cite{shoshitaishvili2015firmalice} and
Karonte~\cite{redini2020karonte} apply handcrafted
symbolic rules to detect authentication bypass and
inter-binary taint in firmware. However, heuristic
methods require the analyst to know in advance what
pattern to look for, they do not generalise to
unexpected bug classes, and they demand substantial
human effort to port across protocols.

\textbf{Statistical-based detection.}
Simpson et al.~\cite{https://doi.org/10.1049/smc2.12083}
compare QUIC against TLS and TCP on a single QUIC
implementation, reporting client delay and server
CPU and memory under attack. We instead compare four QUIC
implementations against each other under the same
attack conditions, and we add a binary view of each
server that lets us interpret the network observation
against the defences the server actually carries.
Specification-level work mechanically checks a formal
model of the QUIC state machine against the published
protocol text~\cite{mcmillan2019quic,crochet2025verifyingquic},
complementing cryptographic analyses that establish
what a conformant implementation should
guarantee~\cite{fischlin2014quic,delignat2021everquic}.
However, statistical and formal approaches require
either a running target or a reference specification,
and neither interrogates the compiled binary that is
actually deployed.

\textbf{Learning-based detection.}
Within protocol analysis,
RFCAudit~\cite{zheng2025rfcaudit} checks protocol C
implementations against their published
specifications at 81.9\,\% precision. On the binary side,
LATTE~\cite{liu2025latte} pairs a disassembler with a
language model to drive taint analysis. However, these
systems either reason over source code and
specifications or target generic vulnerability classes
rather than the protocol-specific binary behaviour we
interpret in light of the network capture. No
prior learning-based study correlates network traffic
statistics with binary inspection across multiple QUIC
implementations under controlled attack conditions,
which is the gap this paper addresses.

\section{Binary and System Integrated Security Analysis (BSISA)}
\label{sec:bsisa}

We assume an attacker that can send arbitrary QUIC
packets to the server from a separate machine on
the same network, including high-rate Initial
floods, slowloris handshake holds, MitM replay,
0-RTT replay, Retry-token forgery, and connection-ID
exhaustion. We do not assume the attacker has
source-code access to the server, can modify the
server's binary, can read the server's keys, or can
break QUIC's cryptography. We assume the defender
has access to the compiled server binary, a packet
capture taken at the server's network interface,
the qlog event log the server emits, and the
system-call trace of the server process. We assume
the binary the defender analyses is the one that
actually runs in production.

We scope this paper to QUIC server implementations
because the six attack scenarios we evaluate all
target server-side defences and the firmware
evaluation use case we target distributes a server
library, not a client. A client binary can in
principle be analysed by the same BSISA approach.

The Binary and System Integrated Security Analysis
(BSISA) has three parts. The first part is a
multi-agent binary analysis we call \emph{Orchestrated
Reasoning for Binary Code Analysis} (ORCA), which
inspects the compiled QUIC server and produces a
per-binary defence inventory keyed by the named QUIC
defences. The second part is
a system-level analysis that consumes the packet
capture and the qlog event log the server emits while
it is running, and produces a runtime feature record.
The third part integrates the two and emits one
attack label per run with a per-defence
justification. Sections~\ref{sec:bsisa:binary}
through \ref{sec:bsisa:integration} describe each
part.

\begin{figure}
\centering
\includegraphics[width=\columnwidth]{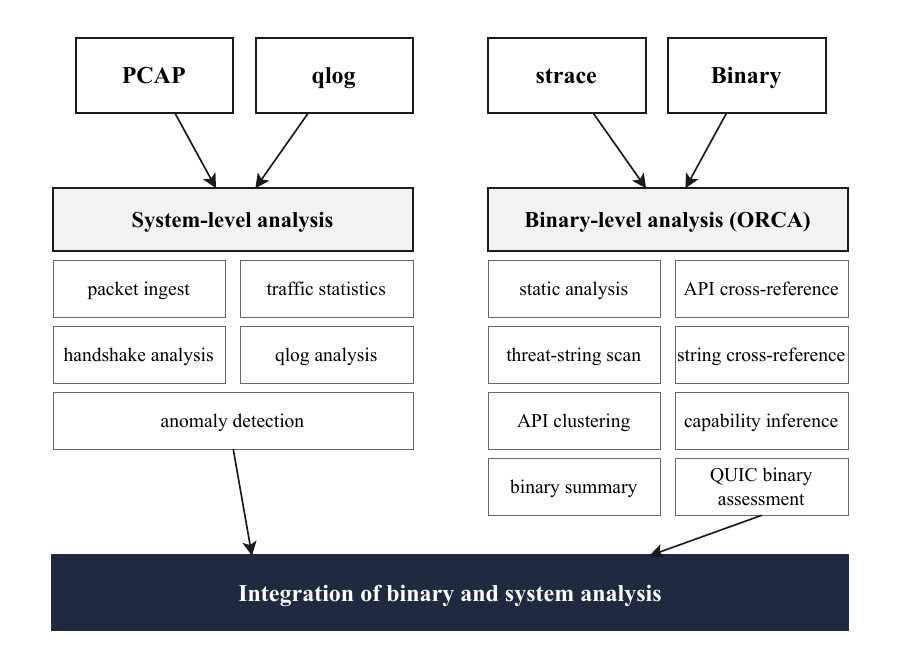}
\vspace{-6pt}
\caption{The structure of the proposed binary- and system-level integrated analysis.}
\label{fig:pipeline}
\end{figure}

\subsection{Orchestrated Reasoning for Binary Code Analysis (ORCA)}
\label{sec:bsisa:binary}

ORCA is a multi-agent binary analysis system for QUIC
servers. The binary analysis comprises five tasks, namely
disassembly of the binary, symbol resolution across
the function table, string analysis over the
read-only data, behavioural clustering of related
functions, and protocol-specific mapping from binary
capabilities to the named QUIC defences. To implement these tasks, we design eight agents
that execute in sequence, each agent's output feeding
the next. Small specialised agents let us inspect
each structured output and locate the stage that
introduced an error when an inference is wrong.

The \emph{static analysis agent} disassembles the
binary and enumerates every function in the symbol
table because some QUIC defences are inlined into
broader packet-processing or timer functions rather
than placed in dedicated routines, and the symbol
table may be stripped in a deployed binary.

The \emph{API cross-reference agent} maps every
imported library function and every system-call name
onto the binary functions that reference it, giving
each function a structural fingerprint based on the
small set of identifiable APIs that most QUIC defences
invoke.

The \emph{threat-string scan agent} scans the
binary's read-only data for strings that QUIC
implementations typically use to name their own
defence routines, such as ``retry'',
``anti\_replay'', ``idle\_timeout'', and
``active\_connection\_id\_limit''. The \emph{string
cross-reference agent} then maps each such string to
the function that loads it. We add the string handle
to the analysis because compiler name-mangling and
symbol stripping can hide the original function name
in the export table, but the string the developer
wrote into the source code is usually still in the
binary's data segment. The string therefore gives us
a textual handle on what a function does even when
the compiled name does not.

The \emph{API clustering agent} groups the APIs each
function calls into a small set of subsystems,
namely TLS, crypto, network input and output, and
timer. Clustering is needed because a single QUIC
defence is usually spread across several functions,
and grouping by the APIs they share lets us see the
defence as one subsystem rather than as a scatter of
unrelated names. The \emph{capability inference
agent} then takes the API clusters, the
function-level string mappings, and the
cross-reference table, and produces a per-binary
capability list, for example ``this binary has
Retry-token validation in subsystem TLS+timer with
three contributing functions''.

The \emph{binary summary agent} compresses the
capability list into a structured summary that
downstream agents can read without re-running the
disassembly. The \emph{QUIC binary assessment agent}
maps the binary summary onto the named QUIC defences
and produces a per-binary defence inventory keyed by
the QUIC defence name. This agent is what connects the protocol-independent
reasoning above to the QUIC-specific evidence the
integration step consumes.

The output ORCA hands to the integration step is a
typed per-defence record, naming the defence, the
protocol section it corresponds to, the binary
function that implements it, and the subsystem it
belongs to.

\subsection{System-Level Analysis}
\label{sec:bsisa:system}

The system-level analysis produces the runtime view
of the server from two sources, namely the packet
capture taken at the server's network interface and
the qlog event log that the server emits while it is
running. We use both rather than only the packet
capture because the QUIC public header is encrypted
under per-connection keys, which means a generic
packet capture cannot read transport-state
information such as packet numbers or stream offsets
without those keys. qlog gives us the transport-state
view from inside the server.

The \emph{packet ingest} agent reassembles QUIC
long-header and short-header packets and keys each
connection by its destination connection identifier.
We key by destination connection identifier rather
than by the IP and port pair because in QUIC the
connection identity is carried in the header rather
than in the network tuple, so a generic packet
sniffer cannot group packets into connections without
protocol-aware reassembly.

The \emph{traffic statistics} agent computes the
connection rate, the single-packet flow ratio, the
unidirectional to bidirectional flow ratio, the mean
number of concurrent flows over the capture window,
and the burst pattern of incoming connections. Each
captures the network signature of a different
generic attack, for example flooding elevates the
connection rate and slowloris elevates the
concurrent-flow mean.

The \emph{handshake analysis} agent inspects the
Initial and Handshake packet-number spaces and
reports the per-connection handshake completion
rate. This is the network-side proxy for whether the
server actually admitted the connection.

The \emph{qlog analysis} agent parses the qlog file
the server emits and counts the events that name
each defence directly, for example Retry-sent,
packet-dropped with an anti-replay reason, and
connection-ID-retired.

The \emph{anomaly detection} agent compares the
cell's traffic and handshake statistics against the
benign-baseline cell for the same implementation,
and emits a per-cell anomaly score.

The output is a runtime feature record that uses the
same per-defence vocabulary as the binary side.

\subsection{Integration of Binary and System Analysis}
\label{sec:bsisa:integration}

The integration is where the binary defence
inventory and the runtime feature record meet. The
binary inventory says ``Retry-token validation is
present''. The runtime record says ``the connection
rate climbed''. Neither, by itself, says
``Retry-token validation was the defence that
absorbed the attack''. The integration draws that
line.

We perform the integration in two layers. The first
layer is per-attack and per-defence. For each attack
scenario that QUIC names a defence
against, we run an attack-specific reasoning agent
that takes six pieces of evidence as input, namely
the binary's symbol table for that defence, the qlog
events that name the defence at runtime, the
system-call counts that show the defence function
executing, the attacker-side ground-truth rate, the
legitimate-client survival rate, and the static
defence inventory from the binary side. The agent
produces a five-field structured record. The fields
are \emph{attack\_confirmed}, indicating whether the
attack was actually delivered to the server,
\emph{binary\_defence\_implementation}, naming the
binary function that implements the relevant
defence, \emph{runtime\_defence\_activation},
indicating whether that defence executed during the
run, \emph{operational\_impact}, summarising the
effect on legitimate clients, and a one-sentence
\emph{summary}. We use the same five-field schema
for all the attack-specific agents because it makes
the per-cell records directly comparable across
attacks and across implementations, and the
structured form is what the second layer consumes.

The second layer is the integrated classifier. It is
the same large-language-model reasoner instantiated
three times over different evidence slices, namely
system-level only, binary-level only, and the union
of both. The reasoner is Anthropic Claude Sonnet 4,
accessed through the Anthropic API at temperature
0.1, with the prompt template released in the
artefacts at the URL in
Section~\ref{sec:eval:attacks}. We hold the reasoner
constant across the
three configurations so that any difference in
classification accuracy can be attributed to the
evidence the reasoner is allowed to see, rather than
to a difference in the reasoner itself. The combined
slice sees the runtime feature record from the
system side, the defence inventory from the binary
side, and the per-attack five-field records from the
first integration layer, and emits one attack label
per cell together with a per-label justification.

The output is one structured record per cell,
listing the implementation, attack scenario, binary
defence inventory, runtime feature record, five-field
records, and three classifier labels with
justifications. Section~\ref{sec:eval} reads this
output for the reported metrics.

\section{Evaluation}
\label{sec:eval}

\subsection{Experimental Platform}
\label{sec:eval:platform}

\begin{figure}[t]
\centering
\includegraphics[width=\columnwidth]{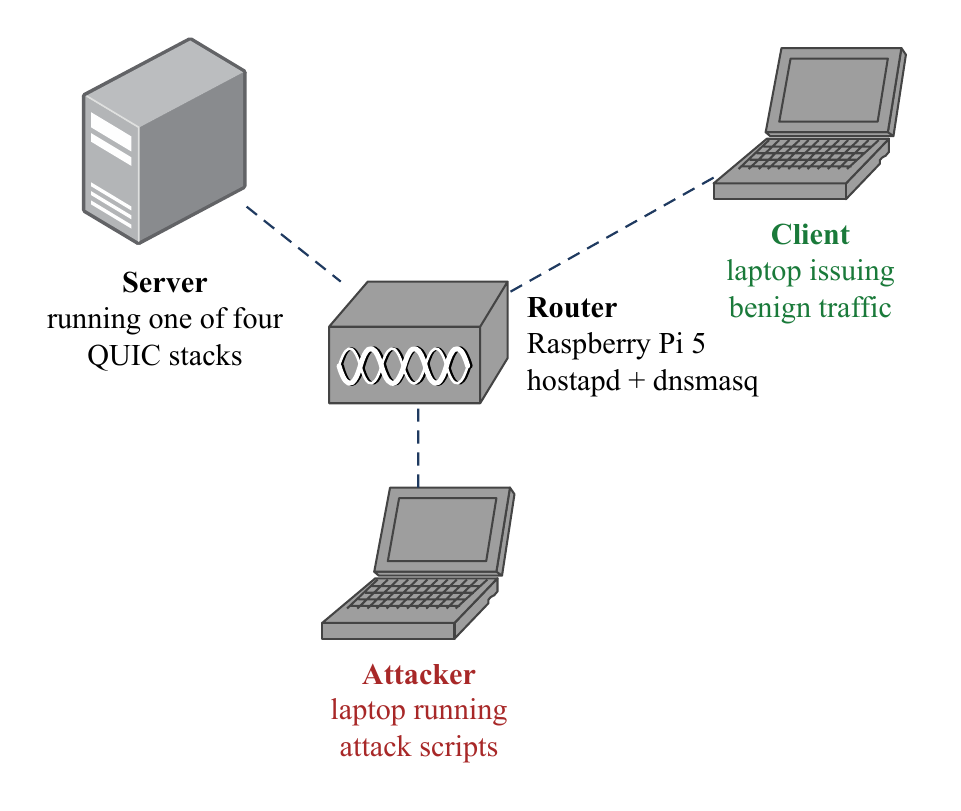}
\vspace{-6pt}
\caption{The network topology used for the experiments.}
\label{fig:topology}
\end{figure}

We evaluate BSISA on a controlled hardware testbed of
four physical machines arranged
as shown in Fig.~\ref{fig:topology}, namely a server
running one of the four QUIC implementations under
study, a client laptop that opens QUIC connections
and exchanges small request-response messages using
the aioquic library~\cite{aioquic}, a router built
from a Raspberry Pi 5 running hostapd as the 802.11
access point~\cite{hostapd} and dnsmasq as the DHCP
server~\cite{dnsmasq}, and an attacker laptop that
runs the attack scripts described in
Section~\ref{sec:eval:attacks}. The router gives the
three endpoints a single shared Wi-Fi network and
isolates the testbed from external traffic so that
bandwidth and delay stay constant across runs.
Table~\ref{tab:testbed} lists the hardware
specifications of the four machines.

\begin{table*}[t]
\centering
\caption{Hardware specifications of the experimental platform.}
\label{tab:testbed}
\renewcommand{\arraystretch}{1.0}
\begin{tabular}{@{}llllll@{}}
\toprule
Device & CPU & No. of threads & Max. speed & Memory & Operating system \\
\midrule
Server   & AMD Ryzen 7 8745HS & 16 & 4.9 GHz & 32 GB & Ubuntu 26.04 \\
Client   & AMD Ryzen 7 4800H  & 16 & 4.2 GHz & 16 GB & Ubuntu 26.04 \\
Attacker & AMD Ryzen 5 3550H  & 8  & 3.7 GHz & 16 GB & Ubuntu 26.04 \\
Router   & ARM Cortex-A76     & 4  & 2.4 GHz & 8 GB  & Raspberry Pi OS \\
\bottomrule
\end{tabular}
\end{table*}

We use four production QUIC server implementations,
namely picoquic 1.1.48.0 (C, picotls) by Christian
Huitema~\cite{huitema-picoquic}, msquic 2.5.5 (C,
quictls) by Microsoft~\cite{microsoft-msquic}, ngtcp2
16.2.3 (C, GnuTLS) by the ngtcp2
project~\cite{ngtcp2}, and quiche 0.28.0 (Rust,
BoringSSL) by Cloudflare~\cite{cloudflare-quiche}.
They are maintained by four independent groups, built
on four different cryptographic libraries, and cover
four deployment settings, namely research and
constrained IoT (picoquic), industrial IoT (msquic),
tooling and library embedding (ngtcp2), and
production HTTP/3 edge (quiche).

The wireless network was dedicated to these four
machines during a run. The server, the client, and
the attacker each kept the same IP address across all
experiments. The server implementation was restarted
at the start of every (implementation, attack) cell
so that state from one cell did not carry over into
the next.

\subsection{Attack Scenarios}
\label{sec:eval:attacks}

Attacks on a secure transport protocol broadly aim
at one of three goals. A first class of attacks
attempts to exhaust the server, and as a result
legitimate connections are degraded or refused. This
class is known as Denial of Service. A second class injects or
tampers with packets on the network path between
the client and the server, which is known as a
Man-in-the-Middle attack. A third class passively
observes the encrypted traffic to learn information
about the client, which is known as traffic
analysis. This paper investigates representative
attacks from the first two classes, because both
can be evaluated against the server's binary
directly. The third class is bound to client-side
traffic patterns and is outside this paper's scope.

For the DoS category, we investigate connection
flooding, slowloris, and connection-ID exhaustion.
Connection flooding is the representative DoS attack
against QUIC's Retry-token and anti-amplification
defences. Slowloris is the representative attack
against QUIC's idle-timeout and connection-cleanup
defences. Connection-ID exhaustion is the
representative attack against QUIC's CID issuance
and retirement defences. For the MitM category, we
investigate packet injection and 0-RTT replay.
Packet injection is the representative attack
against QUIC's AEAD integrity and packet-number
anti-replay defences. 0-RTT replay is the
representative attack against QUIC's 0-RTT
anti-replay defence. Retry-token abuse is the representative attack
against QUIC's Retry-token validation defence, in
which the attacker replays and forges Retry tokens
to bypass address validation. At the rates we
sustained, this attack never triggered Retry mode on
any implementation, so its cells are kept in the
delay sweep of Fig.~\ref{fig:rq2_per_attack} but
excluded from the accuracy calculation in
Section~\ref{sec:eval:rq1}. We additionally run
a benign baseline that gives a reference point for the
server's behaviour with no attack. The attack
implementations and the testbed driver are released
at~\url{https://doi.org/10.5281/zenodo.20338337}.

\subsection{Metrics}
\label{sec:eval:metrics}

\emph{Detection accuracy.} We define detection
accuracy as the fraction of cells where the
classifier's emitted label matches the attack
ground truth. For a
classifier configuration $p \in \{S, B, M\}$, where
$S$ is the system-level configuration, $B$ is the
binary-level configuration, and $M$ is the combined
configuration, and for any set of cells
$\mathcal{X}$ (for example $\mathcal{X}$ is the set
of all 24 cells when reporting overall accuracy, or
the four cells of one attack when reporting
per-attack accuracy), detection accuracy is

\begin{equation}
A(p, \mathcal{X}) = \frac{1}{|\mathcal{X}|}
\sum_{i \in \mathcal{X}}
\mathbf{1}\!\left[\hat{y}_i^{(p)} = G_i\right]
\label{eq:acc}
\end{equation}

where $G_i$ is the ground-truth attack label of cell
$i$, $\hat{y}_i^{(p)}$ is the label emitted by
configuration $p$ on cell $i$, and
$\mathbf{1}[\cdot]$ is the indicator function,
returning 1 when the labels match and 0 otherwise.
The classifier output $\hat{y}_i^{(p)}$ for each
cell is taken as the majority of three independent
samples from the same prompt and configuration, so
that the language model's stochasticity does not
change the result reported here. We use the same
reasoner, the same temperature, and the same
decoding parameters across all 24 cells and all
three configurations, so the variance reported by
the three-sample majority vote is solely due to the
model's sampling and not to changes in
configuration. Detection accuracy is the metric for RQ1, namely
whether the integrated analysis improves accuracy.

We also report \emph{attack-scenario coverage},
defined as the fraction of attack scenarios on which
the configuration correctly classifies at least one
of its four cells. For a configuration $p$ and the
set of attack scenarios $\mathcal{A}$, coverage is

\begin{equation}
C(p) = \frac{1}{|\mathcal{A}|}
\sum_{a \in \mathcal{A}}
\mathbf{1}\!\left[\exists\, i \in \mathcal{X}_a :
\hat{y}_i^{(p)} = a\right]
\label{eq:coverage}
\end{equation}

where $\mathcal{X}_a$ is the set of cells with
attack ground truth $a$. Coverage rewards a
configuration that produces at least some correct
answer on every scenario, even when its cell-level
accuracy is low. Coverage and detection accuracy
together describe the configuration's behaviour
better than either alone, because a configuration
can have high accuracy on a few easy scenarios but
miss every cell of a harder one.

\emph{Mean end-to-end communication delay.} While
each attack runs, a legitimate client on a separate
host repeatedly issues short requests to the QUIC
server. We record, on the client itself, how long
every successful request takes from the moment the
client begins the request to the moment it receives
the server's reply. This is the end-to-end
communication delay observed by the legitimate user
during the attack. Let $\Delta_i$ be the duration of
the $i$-th successful request in one cell. For a
cell that produced $N$ successful requests, the mean
end-to-end communication delay is

\begin{equation}
D = \frac{1}{N} \sum_{i=1}^{N} \Delta_i
\label{eq:delay}
\end{equation}

We compute $D$ from the client-side delay logs of
every cell, and we additionally report the
\emph{legitimate-client failure rate}, namely the
fraction of attempted requests that did not complete,
because an attack that causes the legitimate client
to fail rather than to wait longer is a service
degradation that the mean delay alone does not
capture. Mean end-to-end communication delay and
failure rate together quantify how badly each attack
degrades the legitimate client's service while the
attack is in progress, and how this degradation
differs across the four QUIC server implementations.

\subsection{RQ1, detection accuracy}
\label{sec:eval:rq1}

To answer RQ1 we computed $A(p, \mathcal{X})$ from
Eq.~\ref{eq:acc} for each of the three classifier
configurations on 24 cells. We excluded the
retry-token-abuse cells from the accuracy
calculation because the attack at the rates we ran
never triggered Retry mode on any implementation,
which makes those cells indistinguishable from
slowloris in the evidence record.

The combined configuration is the only one with full
attack-scenario coverage. By Eq.~\ref{eq:coverage},
$C(M) = 6/6$ for the combined configuration,
$C(B) = 5/6$ for the binary-level configuration
because it misses 0-RTT replay, and $C(S) = 5/6$ for
the system-level configuration because it misses
MitM injection. At the cell level the combined
configuration also leads with $A(M) = 45.8\%$, ahead
of $A(B) = 37.5\%$ and $A(S) = 25.0\%$ across the
same 24 cells. The combined view improves on the
system-level view by 20.8 percentage points and on
the binary-level view by 8.3 percentage points. We
report 95\% Wilson confidence intervals on all three
proportions, namely $[12.0\%, 44.9\%]$ for the
system-level configuration, $[21.2\%, 57.3\%]$ for
the binary-level configuration, and $[27.9\%, 64.9\%]$
for the combined configuration.

The Wilson intervals on the binary-level and
combined configurations overlap substantially, so we
do not claim the 8.3 percentage point gain at the
cell level as statistically separated at this sample
size. We instead read the cell-level numbers as a
point estimate and rely on attack-scenario coverage
as the discriminator between configurations, because
coverage is a categorical property of each
configuration's behaviour across the six scenarios
and does not depend on the cell count. By that
measure the combined configuration is the only one
that produces at least one correct cell on every
scenario in our data, which is the property we use
to motivate the combined approach in the rest of the
paper.

\begin{figure}[!t]
  \centering
  \includegraphics[width=\columnwidth]{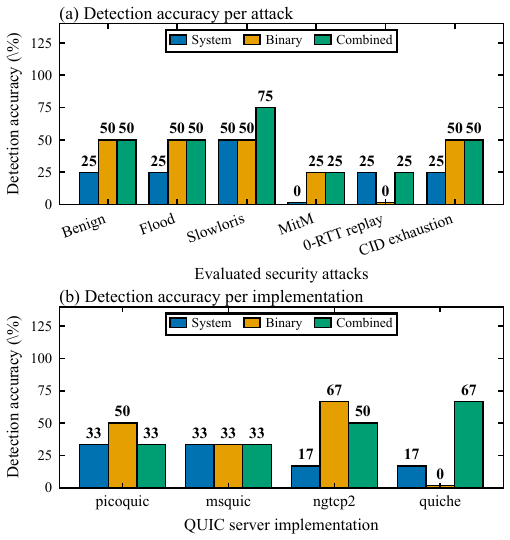}
  \caption{Detection accuracy per attack averaged
  across the four QUIC server implementations (a),
  and per QUIC server implementation averaged across
  the five attack scenarios (b), for the
  system-level, binary-level, and combined classifier
  configurations.}
  \label{fig:rq1_summary}
\end{figure}

Fig.~\ref{fig:rq1_summary} reports the result at two
aggregation levels. Panel~(a) averages each
configuration's per-cell correctness across the four
implementations for each of the five attack
scenarios. Panel~(b) averages across the five attack
scenarios for each implementation. On MitM injection
the system-level configuration is correct on zero
implementations, because the AEAD integrity check at
the server silently rejects the forged packet and no
anomaly is visible in the packet capture statistics.
On 0-RTT replay the binary-level configuration is
correct on zero implementations, because the static
defence inventory of the run under attack is
identical to the inventory of the benign run on every
implementation we tested. On CID exhaustion the binary view alone identifies
the attack on two implementations (picoquic and
ngtcp2). This is a case where the binary view is
already sufficient, and the integration does not
extract additional certainty over the binary view.
The binary deep agent produces a confident
classification because it reads the CID-issuance
and CID-retirement function symbols directly. The system-level view of the same
two cells shows an elevated connection rate. Flood,
CID exhaustion, and slowloris all elevate the
connection rate, so the combined reasoner cannot
disambiguate between them on this evidence and
reverts to the broader FLOODING\_DOS label. The
same effect appears in Fig.~\ref{fig:rq1_summary}(b)
for ngtcp2, where the combined classifier (50\%)
underperforms the binary-only classifier (67\%)
because the elevated connection rate visible to the
system-level slice pulls the reasoner toward a
broader DoS label on cells the binary slice
classifies correctly. This is the cost of adding
low-specificity evidence to a high-specificity
slice.

Beyond the classification accuracy, ORCA's
binary-level inventory exposes a pattern that the
system-level view cannot see. Three of the four
implementations include a Retry-token validation routine
in the binary (\texttt{BIO\_*\_retry\_*} in msquic,
\texttt{Server::verify\_retry\_token} in ngtcp2,
\texttt{quiche\_server::mint\_token} in quiche), yet
the runtime feature record shows no Retry packets
emitted in any of the four flooding cells. ORCA also
finds an anti-amplification routine
(\texttt{amplification\_protection}) in the quiche
binary that does not activate in any cell we ran. We
call these \emph{declared-but-silent} defences. A system-level analysis only sees what the server
does at runtime, so it cannot detect defences that
exist in the binary but never execute. The
binary-level view is what reveals them. The
distinction matters because a defence that never
runs provides no protection regardless of what the
protocol mandates.

The BSISA pipeline analyses each (implementation,
attack) cell in approximately 7.7 minutes on
average (mean 459.6\,s, median 392.2\,s, range
312\,s to 758\,s across the 24 cells). The cost should be
read against manual reverse engineering of the
server binary by a human analyst, which is a
demanding cognitive task with hour-scale per-binary
effort even in controlled empirical
studies~\cite{votipka2020observational,mantovani2022remind}
and which does not scale across multiple
implementations or protocol versions~\cite{caballero2013automatic}.
Because the binary is fixed for a given deployment,
the per-binary cost amortises over every subsequent
traffic capture classified against that binary.

\subsection{RQ2, end-to-end communication efficiency under attack}
\label{sec:eval:rq2}

\begin{figure}[!t]
  \centering
  \includegraphics[width=\columnwidth]{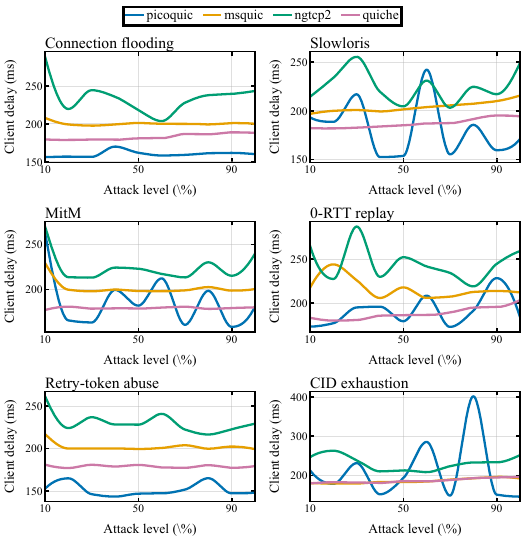}
  \caption{Mean client end-to-end delay per QUIC server
  implementation, swept across attacker rate, for each
  of the six attack scenarios.}
  \label{fig:rq2_per_attack}
\end{figure}

We answer RQ2 by sweeping each attack scenario across
eleven attacker levels and measuring the legitimate
client's end-to-end transaction delay at every level
using Eq.~\ref{eq:delay}. An \emph{attacker level} is
the attacker's send rate expressed as a percentage of
the maximum rate we configured for that attack
scenario. The maximum rate is the highest rate the
attacker laptop sustains for that attack without
dropping packets on its own outbound interface, which
for the flooding and connection-ID exhaustion attacks
is approximately 440 connections per second. Level 0
is no attack, level 50 is half the maximum rate, and
level 100 is the maximum rate.
The eleven levels are spaced in ten-percent steps. At
every level we measure 50 legitimate-client
transactions. We exclude level 0 from
Fig.~\ref{fig:rq2_per_attack} because the
level-0 transactions are taken against a cold server
and are dominated by the one-time handshake cost,
which is not a property of the attack. The plotted
range is level 10 through level 100.
Fig.~\ref{fig:rq2_per_attack} reports the result
across the six attack scenarios. Each panel plots
one scenario with the attacker level on the x-axis,
the mean client delay on the y-axis, and the four
QUIC server implementations as smoothed lines.

Across the six panels of
Fig.~\ref{fig:rq2_per_attack}, the legitimate
client's mean delay stays approximately constant as
the attacker rate increases, on every implementation
and in every attack scenario. quiche records the
lowest delay, between 190~ms and 200~ms across all
six attacks. picoquic and msquic record similar
delay levels, between 170~ms and 245~ms. ngtcp2
records the highest, between 250~ms and 275~ms.
None of the four implementations shows a rising
delay curve in any panel, which means the attacker
traffic at the rates we ran does not measurably slow
the legitimate client's transactions on any
implementation we tested.

On two attack scenarios, the legitimate client on
picoquic does not complete a majority of its
transactions. Under slowloris the failure rate
reaches 72.4\% (398 of 550 attempts failed), and
under connection-ID exhaustion it reaches 73.3\%
(403 of 550 failed). Fig.~\ref{fig:rq2_per_attack}
plots only the transactions that completed, so the
picoquic curves on those two attacks summarise a
small minority of the legitimate-client attempts.
The other three implementations hold their failure
rates at or below 0.5\% across every cell, so their
curves represent the full legitimate-client
experience. For msquic, ngtcp2, and quiche, the six
attacks at the swept rates do not degrade end-to-end
communication efficiency. For picoquic, slowloris
and connection-ID exhaustion degrade efficiency by
denying service to the legitimate client rather than
by slowing it down.

\section{Conclusion}
\label{sec:conclusion}

This paper presented the Binary and System Integrated
Security Analysis (BSISA) approach for QUIC, evaluated
on four production implementations under six attacks
on a hardware testbed. The combined classifier
achieves 45.8\% detection accuracy, ahead of the
binary-level (37.5\%) and system-level (25.0\%)
configurations, and is the only configuration with
full attack-scenario coverage. The legitimate
client's mean end-to-end delay stays approximately
constant under every attack, but picoquic loses
72.4\% of attempted transactions under slowloris and
73.3\% under connection-ID exhaustion, while the
other three implementations hold the failure rate at
or below 0.5\%. The four implementations cover a
range of deployment settings IoT firmware is likely
to encounter, so the findings are informative for
those settings.

BSISA identifies which protocol-level defences each
compiled QUIC server binary implements and which of
them activate under attack. Future work will extend
the binary investigation to identify vulnerabilities
in QUIC implementations, broaden coverage to further
attacks such as version negotiation and connection
migration, and evaluate the approach on
representative IoT hardware such as ESP32-class
gateways running QUIC.

\FloatBarrier

\bibliographystyle{IEEEtran}
\bibliography{biblio_new}

\end{document}